
\input phyzzx.tex

\def\d {\partial}
\def\v {\varphi}
\def\l {\lambda}
\def\a {\alpha}
\def\g {\gamma}
\def\z {\bar z}

\def\f {\phi}

\def\G {\Gamma}
\def\F {\Phi}
\def\picc {\scriptscriptstyle}
\def\s {\sigma}
\def\e {\epsilon}
\def\D {\Delta}
\def\dd {\delta}
\def\L {\Lambda}
\def\ze {\zeta}

\titlepage
\title{The Quantization of Anomalous Gauge Field Theory}

\author {M. Martellini\foot{On leave of absence from Dipartimento di Fisica,
Universit\'a
di Milano, Milano, Italy and I.N.F.N., Sezione di Pavia, Italy}}
\address{I.N.F.N., Sezione di Roma "La Sapienza", Roma, Italy}
\author {M. Spreafico}
\address{Dipartimento di Matematica, Universit\`a di Milano, Milano
and I.N.F.N., Sezione di Milano, Italy}
\author {K. Yoshida}
\address{Dipartimento di Fisica, Universit\`a di Roma, Roma and
I. N. F. N., Sezione di Roma, Italy}
\abstract{We discuss the so called gauge invariant quantization
of anomalous gauge field theory, originally due to Faddeev and Shatashvili.
It is pointed out that the further non invariance of relevant path
integral measures poses a problem when one tries to translate it
to BRST formalism.
The method by which we propose to get around of this problem introduces certain
arbitrariness in the model. We speculate on the possibility of using such
an arbitrariness to build series of non equivalent models of two dimensional
induced gravity.}

\endpage

\pagenumber=1

\vglue 0.6cm
{\bf 1. Introduction}
\vglue 0.4cm
The consistent quantization of (classical) gauge invariant field theory
requires the complete cancellation of anomalies [1] [2]. Here, "consistent"
means that we want not only to require
renormalizability (perturbative finiteness), but also
unitarity of S-matrix, non-violation of Lorentz invariance etc.
Moreover, in physical $4d$ world, anomaly cancellation condition itself
often leads to the physical predictions. The well known example is the equality
of numbers of quarks and leptons in the Standard Model of Weinberg and Salam.

On lower dimensional (eg. $d=2$) field theory, the cancellation of
anomalies is still the crucial ingredient for the model building. The critical
string dimension $d=26$ is often quoted [3] as the consequence of anomaly
free condition for bosonic string (although in this example the cancellation
of anomaly does not guarantee full consistency of the model
in above sense, due to
the presence of tachyons).

In the case of lower dimensional field theory ($d<4$), one often
tries to
quantize a gauge field theory when there is no way of cancelling
its anomaly. The classical example of this situation
is the attempt to the quantization of chiral
Schwinger model by Jackiw and Rajaraman [4] [5]. They have shown that the
model can be consistently quantized (=free field theory) even when the gauge
invariance is broken through anomaly.

In general there seem
to be two ways for attempting the quantization of anomalous
gauge field theory:

\noindent{1) {\it Gauge non invariant method}}

One ignores the breaking of
gauge symmetry and try to show that the theory can be
quantized even without the gauge invariance. The example of this approach
is the above Jackiw-Rajaraman quantization of the chiral Schwinger model. The
problem here is that it is not easy to develop the general technics covering
wide class of physically relevant models with anomaly.

\noindent{2) {\it Gauge invariant method}}

In this case, one first tries to recover gauge invariance
by introducing new degrees of freedom. The theory is anomalous
when one can not find local counter term to cancel the gauge non invariance
due to the one loop "matter" integrals in presence of gauge fields,
by making use
exclusively of the degrees of freedom (fields) already present in the classical
action.

In ref. [6], Faddeev and Shatashvili (FS) have tried to justify the
introduction of new degrees of freedom which are necessary to construct the
anomaly cancelling counter term. Their argument is based on the idea of
projective representation of gauge group.
They observe that the appearance of anomaly does
not mean the simple breakdown of (classical) gauge symmetry,
but it rather signals
that the symmetry is realised projectively (this is related to the appearance
of anomalous commutators of relevant currents). Such a realization,
through projective representations, necessitates the enlargement of
physical Hilbert space. Thus they argued that the introduction
of new fields in the model is not an ad hoc (and largely arbitrary)
construction.

Independently of their "philosophy", the FS method gives the gauge
invariant action at the price of introducing the extra degrees of freedom
(generally physical). The serious problem of this method is, however, that
the gauge invariance thus "forced" upon the theory, does not automatically
guarantee the consistency of the theory. This is in contrast with our
experience with some $4d$ models such as the Standard Model.

For example, one may apply the FS method to the celebrated case of chiral
Schwinger model [4] [5] [5A]. In this case, we have the classical action
$$
S_{\picc 0}= \int {dz\wedge d\z\over 2i}
\Big [\bar\psi_{\picc R}\g_{\z}(\bar\d+R)\psi_{\picc R}+
\bar\psi_{\picc L}\g_z\d\psi_{\picc L}+{1\over 4} Tr~F^2\Big]
$$
where
$$
\eqalign{&\psi_{\picc R/L}={1\pm \g_5\over 2}\psi\cr
&R/L=A_1\pm iA_2, ~F=\bar\d L-\d R+[R,L]\cr}
$$
(we are using the euclidian notation).

So is invariant under the gauge transformation
$$
\eqalign{
&\psi_{\picc R}\to \psi_{\picc R}^g=S(g) \psi_{\picc R}\cr
&\psi_{\picc L}\to\psi_{\picc L}\cr
&A_\mu=g A_\mu g^{-1}+g\d_\mu g^{-1}\cr}
$$
for any $g(z,\z)\in G$.

The theory is anomalous because the one loop integral
$$
e^{-W_{\picc R}(R)}=\int {\cal D}\psi_{\picc R}{\cal D}\bar\psi_{\picc R}
exp-\int\bar\psi_{\picc R}\g_{\z} (\bar\d+R)\psi_{\picc R}
$$
is not gauge invariant under
$$
R\to g R g^{-1}+g\bar\d g^{-1}
$$
(for any choice of the regularization).

Following FS'technic (see next section) however, one can introduce the local
counter term $\L(R, L;g)$, $(g(z,\z)\in G)$ so that the gauge variation
of $\L$ cancels the non invariance of $W_{\picc R}(R)$.

There is certain arbitrariness in the choice of $\L$ but the convenient one
is
$$
\L(R,L;g)=-\Big ( \a_{\picc L}(L,g) + {1\over 4\pi} \int Tr(RL)\Big)
$$
where
$$
\eqalign{
&\a_{\picc L}(L,g)={1\over 4\pi}\Big[ -\int {dz\wedge d\z\over 2i} Tr(g^{-1}
\bar\d g,L)+{1\over 2}\int {dz\wedge d\z\over 2i}
Tr(g\d g^{-1},g\bar\d g^{-1})\cr
&-{1\over 2}\int_0^1dt\int {dz\wedge d\z\over 2i}Tr(g'\d_tg'^{-1},[g'\d
g'^{-1},
g'\bar\d g'^{-1}])\Big]\cr
&g'(0,z,\z)=1,~~~g'(1,z,\z)=g(z,\z)\cr}
$$
is the Wess-Zumino-Novikov-Witten action corresponding to the anomaly of left
fermion $\psi_{\picc L}$, $\bar\psi_{\picc L}$ ($\a_{\picc L}$ is not globally
a local action but it is so for "small" $g\simeq 1+i\xi$).
That is, one can write
$$
\a_{\picc L}(L,g)=W_{\picc L}(L^g)-W_{\picc L}(L)
$$
where
$$
e^{-W_{\picc L}(L)}=\int {\cal D} \psi_{\picc L}'{\cal D}\bar\psi_{\picc L}'
exp-\int \bar\psi_{\picc L}'\g_z (\d+L)\psi_{\picc L}'
$$
(note that $\psi_{\picc L}',\bar\psi_{\picc L}'$ have nothing to do with
$\psi_{\picc L},\bar\psi_{\picc L}$ in $S_{\picc 0}$).

With this choice of counter term, one can show that the theory is equivalent
to {\it a}) free decoupled fermion $\psi_{\picc L},\bar\psi_{\picc L}$ and
{\it b}) the vector Schwinger model. In fact, the added bosonic degree of
freedom $g(z,\z)\in G$ can be "fermionized" to act as missing $\psi_{\picc L}',
\bar\psi_{\picc L}'$ with the
right coupling to the left component $L$ of gauge field.

However, there is still a point missing in this story. In fact, after
introducing the new degree of freedom $g$, there is no reason to exclude
the other type of invariant local counter term such as
$$
{a\over 4\pi}\int Tr (L^g R^g)=
{a\over 4\pi}\int Tr\Big( (g L g^{-1} + g \d g^{-1}), (g R g^{-1}
+g \bar\d g^{-1})\Big)
$$
(one can also attribute it to indefinite - regularization dependent -
part of fermionic integral, i.e. $W_{\picc R}(R)+W_{\picc L}(L) + {a\over
4 \pi}\int Tr (RL)$).

It is well known [4] that the arbitrary constant $a$
enters the physical spectrum.
For abelian case, $G=U(1)$, the mass square of massive boson is given by
$$
m^2={e^2 a^2\over a-1}
$$
thus, for $a<1$, the theory is not consistent although the requirement of gauge
invariance is satisfied.

In the fermionized version of the theory [5A],
$a$ enters the charges of left and right fermions as
$$
e_{\picc R/L}={e\over 2}\Big( \sqrt{a-1}\pm {1\over \sqrt{a-1}}\Big)
$$

This means that the condition $a>1$ is necessary also for the real coupling
constant, or the hermitian hamiltonian.

In general, the consistence of the theory can be proved if one can set
up the BRST scheme with certain physical conditions at the start, such as
hermiticity of the hamiltonian [11].

In what follows, we discuss the possibility of recasting the FS method into
BRST formalism, thus facilitating the analysis of the consistency
of the theory.

\vglue 0.6cm
{\bf 2. Faddeev-Shatashvili method}
\vglue 0.4cm
\noindent{\it a) Path integral formalism}

We shall briefly describe the Faddeev-Shatshvili (FS) method of quantizing
anomalous gauge field theory in the path integral formalism, following
the work of Harada and Tsutsui [7], Babelon, Shaposnik and Vialet [8].

Let us take a generic gauge field theory described by the classical action
$$
S_{\picc 0}(A,X)=S_{\picc G}(A)+S_{\picc M}(X;A)
\eqn\uno
$$
where $\{A(x)\}$ and
$\{X(x)\}$ represent respectively gauge fields and "matter fields",
gauge invariantly coupled to the former.

The total action $S_{\picc 0}$
as well as the pure gauge part $S_{\picc G}$ and the matter
part $S_{\picc M}$ are invariant under the local gauge
transformation
$$
\eqalign{&A\to A'=A^g\cr &X\to X'=X^g\cr}~~~~ g(x)\in G.
\eqn\due
$$

Being anomalous
generally means that the one loop matter integral (assumed that $S_{\picc M}
(X,A)$ is quadratic in  $X$)
$$
\int {\cal D}X e^{-S_{\picc M}(X,A)}\equiv e^{-W(A)}
\eqn\tre
$$
cannot be regularized in such a way as to preserve the gauge invariance
of the functional $W(A)$,
$$
W(A^g)-W(A)=\a(A;g)\not= 0
\eqn\quattro
$$

Naturally, $\a(A;g)$ depends on the regularization used,
but there is no way of cancelling it completely by adding
some local counter
term $\L(A,X)$ to the action.

One can understand eq. \quattro\ as the non invariance of the path
integral measure, ${\cal D} X$:
$$
{\cal D} X^g\not= {\cal D} X
\eqn\cinque
$$

In fact, as shown by Fujikawa [9], one can write the "anomaly equation"
$$
\eqalign{&W(A^g)-W(A)=\a(A;g)\cr
&det\left({{\cal D}X^g\over {\cal D} X}\right) = e^{-\a(A;g)}=
e^{\a(A;g^{-1})}\cr}
\eqn\sei
$$

In this situation, clearly one can not hope that the usual Faddeev-Popov
(FP) ansatz to quantize the theory may go through.

If one inserts the $\delta$ function identity
$$
1=\D(A)\int {\cal D} g \delta(F(A^g))
\eqn\sette
$$
where $F(A)$ is a gauge fixing function, into the path integral
expression for the partition function
$$
Z=\int {\cal D} A\int {\cal D}X e^{-(S_{\picc G}(A)+S_{\picc M}(X;A))}
$$
then one obtains
$$
\eqalign{Z=&\int {\cal D} A \int {\cal D} X \D (A) e^{-[S_{\picc G}(A)+
S_{\picc M}(X;A)]}
\int {\cal D} g \delta (F(A^g))\cr
=&\int {\cal D} A\int {\cal D} g\D (A) e^{-S_{\picc G}(A^g)}
\int {\cal D} X e^{-S_{\picc M}
(X^g;A^g)} \delta (F(A^g))\cr}
\eqn\otto
$$
the second equality follows from
the gauge invariance of the classical action: $S_{\picc 0}(A^g;X^g)=
S_{\picc 0}(A;X)$.

In the case of usual gauge field theory, such as chiral Schwinger model,
we can make a series of
assumptions on the remaining functional measures ${\cal D} A$ and
${\cal D}g$.

First, we assume
$$
(1)~~~~~~~{\cal D} A={\cal D}A^g
\eqn\nove
$$
then with the change of variable $A^g\to A$ and $X^g\to X$ in \sette, we get
$$
Z=\int {\cal D} g  \int {\cal D} A\D \delta(F(A))(A^{g^{-1}})
e^{-S_{\picc G}(A)}
\int {\cal D} X e^{-[S_{\picc M}(X;A)+\a(A;g^{-1})]}
\eqn\dieci
$$
where we have used eq. \cinque, i.e. ${\cal D}X={\cal D}X^{gg^{-1}}=
{\cal D}X^g e^{-\a(A;g^{-1})}$.

Further, one can assume for the usual gauge group,
the invariance of Haar measure ${\cal D} g$, i.e. for any $h$ in $G$
$$
(2)~~~~~~~{\cal D}(gh)={\cal D}(hg)={\cal D} g
\eqn\undici
$$
which results, as is well known, in the invariance of the FP factor
$\D(A)$
$$
\D(A^{g^{-1}})=\D(A)
\eqn\dodici
$$

Thus, we get the expression for $Z$ proposed in ref.s [6] and [7]
$$
Z=\int {\cal D} g \int {\cal D} A\D (A)\delta(F(A))\int {\cal D} X
e^{-S_{\picc eff}(X,A;g)}
\eqn\tredici
$$
with
$$
S_{\picc eff}(X,A;g)=S_{\picc 0}(X,A;g)+\a(A;g^{-1})
\eqn\quattordici
$$

As one can see from eq. \quattro\ the effect of the counter term
$\a(A;g^{-1})$ is to transform the one loop path integral $W(A)$, eq. \tre,
to $W(A^{g^{-1}})$, which is trivially gauge invariant under the extended gauge
transformation
$$
\eqalign{
&A\to A^h,~~~X\to X^h\cr
&g\to hg\cr}
\eqn\quindici
$$
and thus the model is invariant up to one loop level.

We have repeated here the above well known manipulations [7] to emphasize the
relevance of the invariance conditions 1) and 2) (eq.s \nove\ and \undici).

In many familiar example, such as the chiral Schwinger model, these conditions
are trivially satisfied.

One well known case where these conditions
become problematic is the $2d$ induced gravity or off-critical string.
In this case, if one fixes
the path integral measures ${\cal D} \f$ for the Weyl
factor of metric and ${\cal D}\s$ for the Weyl group element
by the invariance under the diffeomorphisms of $2d$ manifold,
then they are not invariant under the translations, eg. $\s\to\s+\a$ (i.e.
Weyl transformation). Thus, the path integral measure (i.e. ${\cal D} A
{\cal D} g$)
can never be invariant under the whole gauge group
$$
G=Diffeo\otimes Weyl
$$

\noindent{\it b) BRST [10] quantization}

A more rigorous strategy to have a consistent formulation of a
gauge field theory is to recast
it in the BRST formalism. In this way, one may discuss the
physically important questions such as the unitariety of S-matrix [11].

In the simpler example like the chiral gauge field theory where the
invariance of the measure
${\cal D} g {\cal D} A$, eq.s \nove\ and \undici,
under the gauge transformations are respected,
there is no difficulty in setting up the BRST procedure once the anomaly
has been removed.

One replaces the "heuristic" FP factor
$$
\D(A)\delta(F(A))=det \left ( {\delta F(A^h)\over \delta h}\Big|_{h=1}
\right) \delta (F(A))
$$
with BRST gauge fixing term
$$
exp-\int\hat s (\bar c F(A))= exp-\int\Big[B F(A)-\bar c {\delta F(A^h)\over
\delta h}\Big|_{h=1} c\Big]
$$
where $c,\bar c$ are the BRST ghosts corresponding to the gauge group $G$
while $B$ ("Lagrange multiplier") is the Nakanishi-Lautrup field.
Under the BRST operator $\hat s$, one has, in particular
$$
\eqalign{&\hat s \bar c= B\cr
&\hat s B = 0\cr
&(\hat s^2=0)\cr}
$$

The counter term $\a(A;g)$ cancelling the one-loop anomaly, one can show
easily the validity of the Slavnov-Taylor identity
$$
\eqalign{
&{\dd \tilde\G\over\dd A}{\dd \tilde\G\over\dd K}+{\dd \tilde\G\over\dd \F_i}
{\dd \tilde\G\over\dd K_i}+{\dd \tilde\G\over\dd c}{\dd \tilde\G\over\dd L}\cr
&(\tilde\G*\tilde\G=0)\cr}
\eqn\sedici
$$
up to one loop.

$\tilde\G$ is the generating functional of the one particle irreducible
part $\G$ (with added external source for composite operators) minus the
"gauge fixing term" (in \sedici, $A$ and $c$ are the classical counter
parts of the gauge fields $A$ and ghost $c$, while $\{\F_i\}$ are the classical
fields for the matter $X$ and newly introduced field $g$; $K,K_i$ and $L$ are
the usual external sources for the gauge variations $\hat\dd A,\hat\dd\F_i$
and $\dd c$ respectively).
One then hopes that it is possible to chose the higher order local counter
term in such a way that eq. \sedici\ is satisfied to all orders.

Let us now imagine, however, that the invariance conditions 1) and 2)
for the measure ${\cal D} A {\cal D} g$
(eq.s \nove\ and \undici) are not satisfied.
This means that one should take account of one or both of the following
situations:

(1') the condition (1) is not satisfied, i.e. ${\cal D} A\not={\cal D} A^g
={\cal D} A e^{-\a'(A;g)}$, where $\a'(A;g)$ is the "Fujikawa determinant"
associated with the non gauge invariance of measure over gauge field
itself.

(2') the condition (2) is not satisfied, i.e. $\D(A^g)\not=\D(A)$.

First of all, the non invariance property 2') means that the factor
$\D (A)\dd (F(A))$ in eq. \undici\ must be replaced by $\D(A^{g^{-1}})
\dd(F(A))$.

Thus, instead of a BRST gauge fixing term \quattordici\ one ends up with
$$
\int \hat s (\bar c F(A)+ln\left( {\D(A^{g^{-1}})\over \D (A)}
\right)
\eqn\diciannove
$$
The trouble is that one can not transform $-ln\D (A)$ into a BRST invariant
local term in the action. In fact, the BRST gauge fixed action would appear
something like
$$
S_{\picc eff}=S_{\picc 0}+\a(A;g^{-1})+\a'(A;g^{-1})+
ln\left( {\D(A^{g^{-1}})\over \D (A)}\right)+\int \hat s (\bar c F(A)
\eqn\venti
$$

The extra one loop term $\a'(A;g)$ does not cause any trouble for the
BRST scheme to work
at least in the example we are interested. One way to push through the
BRST scheme may be to replace eq. \venti\ with
$$
S_{\picc eff}'=S_{\picc 0}+\a(A;g^{-1})+\a'(A;g^{-1})+\int \hat s (\bar c F(A)
\eqn\ventuno
$$

It is likely that the effective action \ventuno\ leads to a consistent BRST
quantization. One may only add that it does not correspond to the
path integral method of ref.s [7] and [8] when $\D(A^g)\not=\D(A)$.

To reconciliate the "path integral" formulation of FS method with BRST
scheme, we propose another possibility.

It must be realized that once the new degree of freedom $g$ is admitted in the
theory then there is no reason to exclude new local counter terms of the
right dimension which are BRST invariant and which may also depend
on $g$. Naturally this will
change the model and its "physics", but nevertheless
it can remain consistent, in so far as the BRST invariance is maintained.

Let us then introduce the following counter term in our theory
$$
\tilde\L_{\picc G}(A,g;c,\bar c,c',\bar c',B)=\Big[ B G(A^{g^{-1}})-
\bar c'{\delta G(A^{g^{-1}h})\over \delta h}\Big|_{h=1} c'\Big]-
\Big[ B G(A)-
\bar c{\delta G(A^{h})\over \delta h}\Big|_{h=1}c \Big]
\eqn\ventidue
$$
where the second pair of "ghosts" $c',\bar c'$ are defined as the BRST singlet
$$
\eqalign{&\hat\delta\bar c'=0\cr
&\hat s c'=0\cr}
\eqn\ventitre
$$
and $G(A)$ is the "pseudo gauge fixing" which is generally different
from $F(A)$.

The first term in $\tilde \L_{\picc G}$ is trivially BRST invariant since all
the fields involved are either gauge invariant by themselves
or appear as invariant
combinations.
The second term, on the other hand, can be written as
$$
\hat s(\bar c G(A))
$$
so it is invariant too.

The effective action now takes the form
$$
S_{\picc eff}=S_{\picc 0}+\a(A;g^{-1})+\a'(A;g^{-1})+
\int\tilde\L_{\picc G}(A,g;c,\bar c,c',\bar c', B)+\int \hat s (\bar c F(A)
\eqn\ventiquattro
$$

Note that the gauge freedom of the BRST invariant theory \ventiquattro\
is represented by the (arbitrary) gauge fixing
function $F(A)$ while each different
choice of "pseudo gauge function" $G(A)$ defines
a new model.

Each choice of $G(A)$ then results in a gauge
invariant model which must then be gauge fixed by choosing a particular form
for $F(A)$.
In the limit of singular gauge
$$
F(A)\to G(A)
\eqn\venticinque
$$
the effective action \ventiquattro\ gives the series of models depending
on $G(A)$ alone. The corresponding effective action can be formally written
$$
S_{\picc eff}=S_{\picc 0}+\a(A;g^{-1})+
\a'(A;g^{-1})+\Big [ B G(A^{^{-1}})-\bar c'
{\delta G(A^{g^{-1}h})\over\delta h}\Big|_{h=1} c'\Big]
\eqn\ventisei
$$

Note that in \ventisei\ the gauge is already fixed (with a singular gauge).
To see the gauge invariance property of the model \ventisei, one must go back
to eq. \ventiquattro\ with \ventidue, i.e.
$$
\eqalign{
S_{\picc eff}^{\picc inv}&=S_{\picc 0}+\a(A;g^{-1})+
\a'(A;g^{-1})+\int\Big [ B G(A^{^{-1}})-\bar c'
{\delta G(A^{g^{-1}h})\over\delta h}\Big|_{h=1} c'\Big]\cr
&-\int\Big [ B [F(A)-G(A)]-\bar c{\delta \over\delta h}[
F(A^h)-G(A^h)]\Big|_{h=1} c\Big]\cr}
\eqn\ventisette
$$

We have seen in this way that the FS method of formulating an anomalous theory
within the path integral formalism apparently generates a series of physically
distinct and BRST invariant gauge fields theories.

We will discuss the possible candidate for such a scenario in the next section.

\vglue0.4cm
{\bf 3. Two dimensional induced gravity}
\vglue0.6cm
In this section we would like to apply the FS method of \S 1 to analyze the
quantization problem of $2d$ gravity [15]
(off critical string) in conformal
gauge [16].
The theory at classical level is defined in term of the Polyakov action
$$
S_{\picc 0}=\sum_{\mu=1}^d\int d^2 x \sqrt{g}g^{ab} \d_a X_\mu \d_b X^\mu
\eqn\ventotto
$$
where $\{X^\mu(x)\}_{\mu=1,d}$ are the bosonic matter fields coupled to
the $2d$ metric $g_{ab}$ (in the string language,
the string is immersed in a $d$-dimensional target
space).

We use euclidian metric and introduce the complex coordinates
$$
\eqalign{
&z=x_1+ix_2\cr
&\z=x_1-ix_2\cr}
$$

The invariant line element can be written as
$$
ds^2=g_{ab}dx^a d x^b= e^\f |dz+\mu d\z|^2
\eqn\ventinove
$$

Thus, one can conveniently parametrize the metric as
$$
\eqalign{
&g_{zz}=\bar\mu e^\f,~~~~g_{\z\z}=\mu e^\f\cr
&g_{z\z}=g_{\z z}={1+\bar\mu\mu\over 2} e^\f\cr}
$$

In term of the parameters $\mu,\bar\mu$ and $\f$ the classical action
\ventotto\ takes the form [15]
$$
S_{\picc 0}=\sum_{\mu=1}^d\int {dz\wedge d\z\over 2i} {(\bar\d-\mu\d)X_\mu
(\d-\bar\mu\bar\d)X^\mu\over 1-\mu\bar\mu}
$$

It is understood that $\mu$ and $\bar\mu$ are constrained by
$$
|\mu|^2<1
$$

The classical action $S_{\picc 0}$ is invariant under the gauge
group $G$ which is the semidirect product of Diffeomorphisms
(general coordinate transformations)
and Weyl transformations. These symmetry groups imply respectively:

\noindent
1) the symmetry under the general coordinate transformation
$$
\eqalign{
&z\to z'=f(z,\z)\cr
&\z\to\z'=\bar f(z,\z)\cr}
\eqn\trenta
$$
where the relevant fields transform as follows
$$
\eqalign{
&X^\mu(z,\z)\to X^{\mu'}(z',\z')=X^\mu(z,\z)~~~(scalar)\cr
&\mu(z,\z)\to\mu'(\z',\z')=-{\bar\d f-\mu\d f\over\bar\d\bar f-\mu\d \bar f}
(z,\z)\cr
&\f(z,\z)\to\f'(z',\z')=\f(z,\z)+
ln
{(\bar\d \bar f-\mu\d \bar f)(\d f-\bar\mu \bar\d f)\over D_f^2 }\cr}
\eqn\trentuno
$$
where
$$
D_f=det\left(\matrix{&\d f&\d\bar f\cr &\bar \d f&\bar\d\bar f\cr}\right)
$$

\noindent
2) The symmetry under the local rescaling of the $2d$ metric
$$
g_{ab}\to e^\s g_{ab}
$$
or in term of the $\mu,\bar\mu$ and $\f$ variables
$$
\mu\to\mu,~~~~\bar\mu\to\bar\mu,~~~~\f\to\f+\s
\eqn\trentadue
$$

It is well known that the theory is anomalous, i.e. one can not regularize the
path integral in a way that conserves the whole $G=Diffeo\times
Weyl$ group.

One can see this easily, examining the matter integral measure ${\cal D}X^\mu$.
With the simplest (translationally invariant or "flat") regularization
${\cal D}_{\picc 0}X^\mu$, one has
$$
\prod_{\mu=1}^d\int {\cal D}_{\picc 0} X^\mu e^{S_{\picc 0}(X,\mu,\bar\mu)}
=exp-{d\over 24\pi}[W(\mu)+\bar W(\bar\mu)]
\eqn\trentatre
$$
where $W(\mu)$ is the Polyakov's "light cone gauge" action [13].

This is naturally Weyl invariant ($S_{\picc 0}$ does not contain the variable
$\f$). On the other hand, it is equally clear that one has lost
diffeomorphism's invariance, since the invariance under general
coordinate transformations means
$$
\dd W(\mu)=0
\eqn\trentaquattro
$$
under $\dd\mu=(\bar\d-\mu\d+\d\mu)(\e+\mu\bar\e)$, which corresponds to
the infinitesimal version of eq.s \trentuno\ with
$f(z,\z)=\e(z,\z),\bar f(z,\z)=\bar\e(z,\z)$.

Eq. \trentaquattro\ is equivalent to the functional differential equation
$$
(\bar\d-\mu\d-2\d\mu){\dd W\over \dd\mu(z,\z)}=0
$$

A well known computation [16] gives, instead,
$$
(\bar\d-\mu\d-2\d\mu){\dd W\over \dd\mu(z,\z)}=\d^3\mu\not= 0
\eqn\trentacinque
$$

Thus, ${\cal D}_{\picc 0} X^\mu$ can not be invariant under diffeomorphisms.
One can define the diffeomorphisms invariant measure
${\cal D}_{\picc Diffeo} X^\mu$ by introducing the local counter term
$$
\eqalign{
\L(\mu,\bar\mu,\f)&=-{1\over 2}\int {dz\wedge d\z\over 2\pi} \Big[
{1\over 1-\mu\bar\mu}[(\d-\bar\mu\bar\d)\f(\bar\d-\mu\d)\f\cr
&-2(\bar\d\bar\mu(\bar\d-\mu\d)+\d\mu(\d-\bar\mu\bar\d))\f ]
+F(\mu,\bar\mu)\Big ]\cr}
\eqn\trentasei
$$
where $F(\mu,\bar\mu)$ is a local function of $\mu$ and $\bar\mu$
only. We do not need the explicit form of $F$ [17].

The new effective action
$$
W_{\picc cov}(\mu,\bar\mu,\f)=W(\mu)+\bar W(\bar\mu)+\L(\mu,\bar\mu,\f)
$$
is invariant under diffeomorphisms.

One can write $W_{\picc cov}(\mu,\bar\mu,\f)$ compactly in the form
$$
W_{\picc cov}(\mu,\bar\mu,\f)=\int {dz\wedge d\z\over 2\pi}
{(\d-\bar\mu\bar\d)\F (\bar\d-\mu\d)\F\over 1-\mu\bar\mu}=
\int d^2 x \sqrt{g} g^{ab}\d_a\F\d_b\F
\eqn\trentasette
$$
where $\F=\f-ln\d\ze\bar\d\ze$ and $\mu={\bar\d\ze\over\d\ze}$
(Beltrami differentials).
Non local (with respect to $\mu$ and $\bar\mu$) parameter $\ze(z,\z)$
is Polyakov meson field (13) in $2d$ gravity.

One characterizes the diffeomorphisms invariant measure ${\cal D}_{\picc
diffeo}
X^\mu$ by
$$
\prod_{\mu=1}^d \int {\cal D}_{\picc diffeo}X^\mu e^{S_{\picc
0}(X,\mu,\bar\mu)}
=exp-{d\over 24\pi} W_{\picc cov}(\mu,\bar\mu,\f)
\eqn\trentotto
$$

(One can understand the appearance of $\f$ field, which is absent in the
classical action, as due to the introduction of a
covariant regularization:
$\L_{\picc cov}$, $ds^2\sim e^\f|dz|^2>\L^2_{\picc cov}$).

Following for instance DDK [14], in what follows we consistently make
use of the diffeomorphisms invariant measure.
Thus, except when indicated explicitly otherwise,
$$
{\cal D} X^\mu\equiv {\cal D}_{\picc Diffeo} X^\mu
\eqn\trentanove
$$
and more generally ${\cal D} \v\equiv {\cal D}_{\picc Diffeo}\v$ for any other
filed $\v$.

Evidently, the diffeomorphisms invariant measure ${\cal D} X^\mu$ can not
be invariant under the Weyl transformation
$$
\f\to\f+\s
$$

Thus, one establishes that the theory is $G$ anomalous.

(Faddeev-Shatashvili method)

Having seen that our model for $2d$ gravity is anomalous, one would like to
apply to it
the FS method of "gauge invariant" quantization of \S 1. As in \S 1,
we "preestablish" the gauge choice for the full group $G=Diffeo\times Weyl$
$$
\eqalign{
&\mu=\mu_0\cr
&\bar\mu=\bar\mu_0~~~diffeomorphisms\cr
&F(\f)=0~~~~Weyl\cr}
\eqn\quaranta
$$

Since our regularization preserves the diffeomorphisms we assume that
the gauge fixing problem (with relevant "$b,c$" ghosts) for diffeomorphisms
has been already taken care for.

To deal with anomalous Weyl symmetry, we have to introduce an extra
degree of freedom, a scalar field $\s(z,\z)$, corresponding to the element
of Weyl symmetry group $g=e^{\s(z,\z)}$.

The anomaly cancelling counter term suggested by FS is then given by
$$
\eqalign{
\a(\mu,\bar\mu,\f;-\s)&=W_{\picc cov}((\mu,\bar\mu,\f-\s)-W_{\picc cov}
(\mu,\bar\mu,\f)=\cr
&=-{1\over 2}\int {dz\wedge d\z\over 2i} {1\over 1-\mu\bar\mu}
[(\d-\bar\mu\bar\d)\s(\bar\d-\mu\d)\s+2(\d-\bar\mu\bar\d)\s(\bar\d-\mu\d)\f\cr
&-2(\bar\d\bar\mu(\bar\d-\mu\d)+\d\mu(\d-\bar\mu\bar\d))\f]\cr}
\eqn\quarantuno
$$

Note that the non local part of $W_{\picc cov}$ is cancelled and
$\a(\mu,\bar\mu,\f;-\s)$ is perfectly local. Naturally, one needs the counter
term $\a$ for each covariant one loop integral corresponding not only to the
matter field $\{X^\mu\}_{\mu=1}^d$, but also to the diffeomorphism ghosts,
$b,c$ and $\bar b,\bar c$, as well as to the $\f$ field contained in
$W_{\picc cov}(\mu,\bar\mu,\f)$.

Thus, the effective action in sense of \S 2 is given by
$$
S_{\picc eff}=S_{\picc 0}(X, \mu,\bar\mu)+S_{\picc gf}^{(d)}(b,c,\bar b,\bar c,
B,\bar B, \mu,\bar \mu)+\g'\a(\mu,\bar\mu,\f;-\s)
\eqn\quarantadue
$$
where $S_{\picc gf}^{(d)}$ is the gauge fixing term with respect to the
non anomalous diffeomorphism symmetry.

As explained above, the coefficient $\g'$ is contributed by all the relevant
fields, that is
$\{X^\mu\}_{\mu=1}^d\Rightarrow d, (b,c,\bar b,\bar c)\Rightarrow
-26, \f\Rightarrow 1$, which gives $\g'={d-26+1\over 24\pi}={d-25\over 24\pi}$.

Note that the contribution of $\f$ field is due to the fact that
${\cal D}_{\picc Diffeo}\f\not= {\cal D}_{\picc 0}\f$, or
in the terminology of \S 2, that one needs the "second" FS counter term
"$\a'(\f;\s)$".

One can now write down the partition function $Z$ with the FS prescription
(within the path integral formalism of ref. [7], see eq. \otto\ of
\S 1). Integrating out the "matter fields" ($X^\mu,b,c,\bar b,\bar c, $),
one has
$$
\eqalign{
Z\sim&\int{\cal D}\s{\cal D}\f \Big [ exp -\g'\int {dz\wedge d\z\over 2i}
{1\over 1-\mu_0\bar\mu_0}\big(
(\d-\bar\mu_0\bar\d)(\f-\s)(\bar\d-\mu_0\d)(\f-\s)\cr
&-2(\bar\d\bar\mu_0(\bar\d-\mu_0\d)+\d\mu_0(\d-\bar\mu_0\bar\d))(\f-\s)
\big)\Big]\D(\f-\s)\dd(F(\f))\cr}
\eqn\quarantatre
$$
where the local action in the exponential is essentially a Liouville
action $S_{\picc L}'(\f')$, ($\f'=\f-\s$). The last two factors come from
the $\dd$ function insertion
$$
\D(\f)\int{\cal D}\s\dd(F(\f+\s))=1
\eqn\quarantaquattro
$$

Note that, since ${\cal D}\s\equiv{\cal D}_{\picc Diffeo}\s\not=
{\cal D}_{\picc 0}\s$ (${\cal D}_{\picc 0}\s$ "flat" measure)
$$
\D(\f-\s)\not=\D(\f)
\eqn\quarantacinque
$$

Formally, one can write the $\D(\f-\s)$ factor as a local action with
the help of the "Weyl ghosts" $\psi$ and $\bar\psi$
$$
\D(\f-\s)=\int{\cal D}\psi{\cal D}\bar\psi exp-\int \bar\psi{\dd F(\f-\s)
\over\dd\f}\psi
\eqn\quarantasei
$$

(BRST procedure)

The path integral argument of \S 1 is at best heuristic. It may suggest the
possible models but one can not prove in this way their consistency.

As it has been argued in \S 1, one may start a more precise discussion
after setting up the BRST quantization procedure. The BRST properties of
the type of models we are dealing with here, have been studied in details
for the critical case, i.e. for $d=26$, where the theory is not
anomalous. In ref. [15], the BRST transformation properties of the fields
are given. They may be used to study our (off critical) model.

One has (see eq. \trentuno)
$$
\eqalign{
&\hat\dd X^\mu=(\xi\cdot\d)X^\mu\cr
&\hat\dd \mu=(\bar\d-\mu\d+\d\mu)c\cr
&\hat\dd\f=\psi+(\xi\d)\f+(\d\xi)+\mu\d\bar\xi+\bar\mu\bar\d\xi\cr
&\hat\dd\xi=(\xi\cdot\d)\xi\cr
&\hat\dd c= c\d c\cr
&\hat\dd\psi=(\xi\cdot\d)\psi\cr}
\eqn\quarantasette
$$
where $\xi\cdot\d$ means $\xi\d+\bar\xi\bar\d$.

Here $\hat\dd$ stands for the both Weyl and diffeomorphism symmetries.
The diffeomorphism ghosts $c,\bar c$ are related to the original
($\xi,\bar\xi$) (corresponding to $\dd z=\e(z,\z),\dd\z=\bar\e(z,\z)$)
by
$$
\eqalign{
&c=\xi+\mu\bar\xi\cr
&\bar c=\bar\xi+\bar\mu\xi\cr}
\eqn\quarantotto
$$

To eq. \quarantasette, we must add the transformation of the auxiliary
field $\s(z,\z)$. Since $\s$ must be a scalar with respect to diffeomorphisms
one has
$$
\hat\dd\s=\psi+(\xi\cdot\d)\s
\eqn\quarantanove
$$

Together with the formulae in eq.s \quarantasette\ to \quarantanove,
one consistently finds
$$
\hat\dd^2=0
\eqn\cinquanta
$$

One should add also the diffeomorphisms anti ghost ($b,\bar b$) and Weyl
anti ghost $\bar\psi$ with the corresponding Nakanishi-Lantrup fields
$B$ and $D$. Their transformation properties are
$$
\eqalign{
&\hat sb=B,~\hat s \bar b=\bar B,~\hat s\bar\psi= D\cr
&\hat s B=\hat s \bar B=\hat s D=0\cr}
\eqn\cinquantuno
$$

We have seen, however, that the Faddeev-Popov factor $\D(\f)$ is not Weyl
invariant \quarantacinque. Thus, according to the result of \S 1, one needs
to correct the effective action $S_{\picc eff}$ by modifying the factor
$\D(\f-\s)\dd(F(\f))$ into a BRST gauge fixing term. As we have seen in \S 1,
such a prescription is not unique. Formally, any action of the form
$$
BRST(invariant)+\hat s(\psi F(\f))(BRST exact)
$$
will do the job.

Now the factor $\D(\f-\s)\dd(F(\f))$ can be rewritten in the form
$$
exp-\int\Big( {\cal D}F(\f)+\bar\psi'{\dd F\over\dd \f}(\f-\s)\psi'\Big)
$$

Thus, in order to follow this expression as close as possible, we suggest
to add a counter term of the form of eq. \ventidue\ in \S 1
$$
\tilde\L_{\picc G}(\f,\s;\psi,\bar \psi,\psi',\bar \psi',D)=
\Big[ D G(\f-\s)+\bar\psi'{\dd G\over\dd\f}(\f-\s)\psi'\Big]-
\Big[DG(\f)+\bar\psi{\dd G\over\dd\f}(\f)\psi\Big]
\eqn\cinquantadue
$$
where we have introduced the function $G(\f)$ to distinguish it from
the true gauge fixing term $s(\bar\psi F(\f))$. The new fields $\psi'$ and
$\bar\psi'$ in eq. \cinquantadue\ ($c'$ and $\bar c'$ in eq. \ventidue\ )
are Weyl singlet and transform as
$$
\eqalign{
&\hat\dd\bar\psi'=0\cr
&\hat\dd\psi'=(\xi\cdot\d)\psi'\cr}
\eqn\cinquantatre
$$

With the addition of the counter term $\tilde\L_{\picc G}$, the effective
action now reads
$$
\eqalign{
\tilde S_{\picc eff}&=S_{\picc L}''(\f-\s)+\int \tilde\L_{\picc G}(\f,\s;\psi,
\bar\psi,\psi',\bar\psi',D)+\int \hat s (\bar\psi F(\f))\cr
&=S_{\picc L}''(\f-\s)+
\int\Big[DG(\f-\s)+\bar\psi'{\dd G\over\dd\f}(\f-\s)\psi'\Big]
+\int\hat s(\bar\psi(F-G)(\f))\cr}
\eqn\cinquantaquattro
$$

The expression for $\tilde S_{\picc eff}$ contains two arbitrary
functions $F(\f)$ and $G(\f)$. Their roles are completely different. While
$F(\f))$ is a genuine gauge fixing function, each choice of $G(\f)$
actually defines a new model.

Naturally, the "series" of models (at arbitrary gauge) includes the familiar
cases. For example, if one fix the model by choosing
$$
G=0
$$
one reproduces the physically equivalent formulations of DDK model.

Alternatively, for any given $G$, one may consider the singular gauge limit
$$
F\to G
$$

In this limit the model formally corresponds to the action
$$
\tilde S_{\picc eff}=S_{\picc L}''(\f-\s)+
\int\Big[DG(\f-\s)+\bar\psi'{\dd G\over\dd\f}(\f-\s)\psi'\Big]
\eqn\cinquantasette
$$

This is the type of model treated in ref. [18]. One may further add the
BRST invariant term $-{\l\over 2}\int D^2$ and transform $\tilde S_{\picc eff}$
into
$$
\tilde S_{\picc eff}'=S_{\picc L}''(\f-\s)+
\int\Big[{1\over 2\l}G^2(\f-\s)+\bar\psi'{\dd G\over\dd\f}(\f-\s)\psi'\Big]
\eqn\cinquantotto
$$

Eq. \cinquantasette\ (or \cinquantotto) seems to be the closest BRST
quantizable approximation to the consequence of FS prescription, i.e. the
insertion
$$
1=\D(\f)\int{\cal D}\s \dd(G(\f+\s))
\eqn\cinquantanove
$$

In ref. [18], and in some later works, the choice
$$
G(\f)=R(\f)-R_0
\eqn\sessanta
$$
with $R$ the scalar curvature, has been made. Using \sessanta, the effective
action \cinquantotto\ becomes
$$
\tilde S_{\picc eff}'(\f'=\f-\s),\psi',\bar\psi,)=
S_{\picc L}''(\f')+
\int\Big[{1\over 2\l}(R(\f')-R_0)^2(\f-\s)+
\bar\psi'{\dd R\over\dd\f}(\f-\s)\psi'\Big]
\eqn\sessantuno
$$

Note that the model defined by \sessantuno\ is fully interacting. In particular
{\it a}) the presence of propagating $\psi'$ and $\bar\psi'$ fields and
{\it b}), more importantly, the presence of $\psi', \bar\psi'$ and
$\f'$ (Yukawa)
interaction in \sessantuno, change the parameters in the Liouville type action
$S_{\picc L}''(\f')$. Such a change, which affects the low energy dynamics
of \sessantuno, can not be calculated exactly. It is not easy even to develop
a systematic perturbation expansion [20].
We believe [18] [19] that the modification
represented by eq. \sessantuno\ may result in deviations from the classical
DDK result, when one uses \sessantuno\ to calculate such physical quantities
as string tension and anomalous dimension.

Lastly, it must be mentioned that the BRST invariant term
$$
\int\bar\psi'{\dd G\over\dd\f}(\f-\s)\psi'
\eqn\sessantadue
$$
in \cinquantasette\ could also be obtained from the alternative gauge fixing
$$
S_{\picc gf}=\int \hat s\Big [\bar\psi{\dd G\over\dd\f}(\f-\s)\s\Big]
\eqn\sessantatre
$$

In this case, one can dispense with the extra BRST invariant (for Weyl
transformation) $\psi'$ and $\bar\psi'$ degrees of freedom. The gauge
fixing function is
$$
F(\f,\s)={\dd G(\f-\s)\over\dd\f}(\f-\s)\s
\eqn\sessantaquattro
$$

It looks as if this model is gauge equivalent
to the DDK model, since the gauge choice $G(\f)=\f$ gives the effective
action
$$
S_{\picc eff}=S_{\picc L}'(\f-\s)+\int(\psi\bar\psi+D\s)\sim S_{\picc L}'(\f)
{}~~~~(\s=0)
\eqn\sessantacinque
$$

The Liouville action $S_{\picc L}'$ here is identical to eq. \quarantaquattro\
without further renormalization (eq. \sessantuno\ is a free field action).

\vglue0.4cm
{\bf 4. Conclusion}
\vglue0.6cm

In this note, we have tried to analyze further consequences of the
Faddeev-Shatshvili method of quantizing anomalous gauge fields theories.

In contrast with other authors [5A], we did not try to
show the equivalence with the "gauge non invariant" method of which
the Jackiw-Rajaraman treatment of the chiral Schwinger model is
 a distinguished example. On the contrary, we have argued that, in certain
cases of physical interest, the FS method can be used to generates new models.

The series of "new" $2d$ gravity models proposed here includes the
models in ref.s [18] [19] as well as the Kawai-Nakayama type $(R-R_0)^2$
(or $R^2$) models [21] [22].

To see if the possibility of enlarging in this way the $2d$ (induced) gravity
models really throws some light on the famous problem of the $d=1$ barrier
in $2d$ gravity, we need a more thorough analysis of the consistency
of these models as well as a better understanding of their physical
consequences.

\vglue 0.6cm
{\bf Acknowledgment}
\vglue0.4cm
This work has been completed while one of the authors (KY) stayed at
National Laboratory of High Energy Physics, Tsukuba, Japan (KEK).
It is a pleasure to thank prof. H. Sugawara and M. Ishibashi for the
hospitality. KY acknowledges stimulating discussions
with many members of KEK, in particular: H. Kawai, S. Aoki, T. Yukawa and
M. Ishibashi. The constructive comments from K. Fujikawa, N. Nakazawa and
K. Ogawa are also gratefully acknowledged. The authors thank G.C. Rossi
for the thorough reading of the manuscript. The work is partially supported
by INFN and Italian Minister of Science and University, MPI 40\%.

\vglue 0.6cm
{\bf REFERENCES}
\vglue 0.4cm
\parindent=0.pt

{[1] C. Bouchia, J. Iliopoulos and Ph. Mayer, Phys. Lett. B38 (1972) 519;}

{[2] D. Gross and R. Jackiw, Phys. Rev. D6 (1972) 477;}

{[3] M. Kato and K. Ogawa, Nucl. Phys. B212 (1983) 443;}

{[4] R. Jackiw and R. Rajaraman, Phys. Rev. Lett. 54 (1985) 1219;}

{[5] R. Rajaraman, Phys. Lett. B154 (1985) 305;}

{[5A] There are large number of works accumulated on this subject.
We list below a few of them which seem to be relevant for the present
discussion. The list is by no means complete however.}

{H.O. Girotti, H.J. Rothe and K.D. Rothe, Phys. Rev. D34 (1986) 592;}

{I.G. Halliday, E. Rabinovici, A. Schwimmer and M. Chanowitz, Nucl. Phys.
B268 (1986) 413;}

{D. Boyanovsky, Nucl. Phys. B294 (1987) 223;}

{L. Caneschi and V. Montalbano, Pisa Preprint IFUP Th-20/86 (unpublished);}

{C. Pittori and M. Testa, Zeitschrift fuer Physik C (Particles and Fields)
47 (1990) 487;}

{[6] L.D. Faddeev and S.L. Shatshvili, Phys. Lett. B167 (1986) 225;
L.D. Faddeev, Nuffield Workshop Proc. (1985);}

{[7] K. Harada and I. Tsutsui, Phys. Lett. B183 (1987) 311;}

{[8] O. Babelon, F.A. Shaposnik and C.M. Vialet, Phys. Lett. B177 (1986)
385;}

{[9] K. Fujikawa, Phys. Rev. Lett. 42 (1979) 11195;}

{[10] C. Becchi, A. Rouet and R. Stora, Ann. of Phys. 98 (1976) 287;
C. Becchi, A. Rouet and R. Stora, {\it Comments on Gauge Fixing II};}

{[11] T. Kugo and I. Ojima, Suppl. Prog. Theor. Phys. 66 (1979) 1;}

{[12] N. Nakanishi, Prog. Theor. Phys. 35 (1966) 1111;
B. Lautrup, Kgl. Danske. Videnskab. Selskab., Mat. Fis. Medd. 35 (1967) 1;}

{[13] A. M. Polyakov, Mod. Phys. Lett. A2 (1987) 893; V. G. Knizhnik, A. M.
Polyakov and A. B. Zamolodchikov, Mod. Phys. Lett. A3 (1988) 819;}

{[14] J. Distler and H. Kawai, Nucl. Phys. B321 (1989) 509;
F. David, Mod Phys Lett. A3 (1988) 1651}

{[15] L. Baulieu, C. Becchi and R. Stora, Phys. Lett. B180 (1986) 55;
L. Baulieu and M. Bellon, Phys. Lett. B196 (1987) 142;
C. Becchi, Nuc. Phys. B304 (1988) 513; see also K. Fujikawa, Nucl. Phys.
B291 (1987) 583; K. Fujikawa, T. Inagaki and H. Suzuki, Phys. Lett. B213
(1988) 279;}

{[16] L. Alvarez-Gaum\`e and E. Witten, NUcl. Phys. B234 (1984) 269;}

{[17] M. N. Sanielvici, G. W. Senenoff and Y. Shi-Wu, Phys. Rev. Lett. 60
(1988)
2571}

{[18] M. Martellini, M. Spreafico and K. Yoshida, Mod. Phys. Lett. A7 (1992)
1281; M. Martellini, M. Spreafico and K. Yoshida, Proc. International Workshop
of String Theory (21/26-9-1992) Accademia dei Lincei, Roma;}

{[19] M. Martellini, M. Spreafico and K. Yoshida, Mod. Phys. Lett. A9 (1994)
2009; M. Martellini, M. Spreafico and K. Yoshida, Rome Preprint 5/1994;}

{[20] H. Kawai, Y. Kitazawa and M. Ninomiya, Nucl. Phys. B393 (1993) 280;
H. Kawai, Y. Kitazawa and M. Ninomiya, Nucl. Phys. B404 (1993) 684;}

{[21] H. Kawai and R. Nakayama, Phys. Lett. B306 (1993) 224;}

{[22] T. Burwick, Nucl. Phys. B418 (1994) 257;}

{[23] J. Cohn and V. Periwal, Phys. Lett. B270 (1991) 18.}

\vfill
\endpage
\end